\documentclass[prb,twocolumn]{revtex4}
\usepackage[dvips]{graphicx}
\usepackage{latexsym,amsmath,amssymb,bm,euscript,eufrak}

\def\etal{~\textit{et~al.}} 
\def\ra{\rangle} 
\def\la{\langle} 
\def\rara{\ra\!\ra}
\def\lala{\la\!\la}
\def\Hc{{\rm h.c.}}
\def\rt3rt3{{\sqrt{3} \times \sqrt{3}}}
\def\nacoo{\rm{Na$_x$CoO$_2$ }}

\begin{document}

\title{Possible effects of charge frustration in Na$_x$CoO$_2$:
bandwidth suppression, charge orders and resurrected 
RVB superconductivity}

\author{O. I. Motrunich and Patrick A. Lee}
\affiliation{Department of Physics, Massachusetts Institute of
Technology, Cambridge MA 02139}

\date{October 16, 2003}

\begin{abstract}
Charge frustration due to further neighbor Coulomb repulsion
can have dramatic effects on the electronic properties of
Na$_x$CoO$_2$ in the full doping range.
It can significantly reduce the effective mobility of the charge
carriers, leading to a low degeneracy temperature
$\epsilon_F \lesssim T$.
Such strongly renormalized Fermi liquid has rather unusual
properties---from the point of view of the ordinary metals with
$\epsilon_F \gg T$---but similar to the properties that are actually 
observed in the \nacoo system.
For example, we show that the anomalous thermopower and Hall effect 
observed in Na$_{0.7}$CoO$_2$ may be interpreted along these lines.
If the repulsion is strong, it can also lead to charge order;
nevertheless, away from the commensurate dopings, the configurational
constraints allow some mobility for the charge carriers, i.e.,
there remains some ``metallic'' component.
Finally, the particularly strong bandwidth suppression around
the commensurate $x=1/3$ can help resurrect the RVB superconductivity,
which would otherwise not be expected near this high doping.
These suggestions are demonstrated specifically for
a $tJ$-like model with an additional nearest neighbor repulsion.
\end{abstract}

\maketitle

\section{Introduction}
Recent discovery\cite{Takada} and confirmation\cite{Foo, Schaak, Chou}
of superconductivity in Na$_{0.35}$CoO$_2 \cdot 1.3$H$_2$O has
stimulated many studies of this material, and in particular of its
unhydrated precursor host material Na$_{0.7}$CoO$_2$.
The \nacoo series has been known for more than
five years\cite{Terasaki, Ando, Koshibae, Ray} for its unusual transport
properties such as large thermoelectric power and nearly
linear-$T$ dependence of the resistivity, indicative of
strong correlation effects.  This has been brought up even
further by a number of recent careful
experiments.\cite{WangQ, WangRH, Has, Gav, Lem, Bru}

It appears from these studies\cite{Chou,Ando} that there is a large 
suppression of the valence band width (equivalently, large effective 
mass enhancement)---by an order of magnitude compared with the LDA 
band structure calculations.\cite{Singh}

We suggest that such large renormalization may be caused by
strong Coulomb repulsion between charge carriers on neighboring
Co sites, and point out that a number of unusual properties of
the system may be explained by the resulting low fermion
degeneracy temperature.
We also consider other possible effects of such repulsion,
in particular, charge ordering.

\vskip 2mm
The plan of the paper is as follows.
To be specific, we consider a $tJ$-like model with
additional strong nearest neighbor repulsion $V$.
In sections~\ref{sec:tVmodel}-\ref{sec:VMC}, we concentrate on the 
dominant $t,V$ energetics.  The study is done by
considering Gutzwiller-like trial fermionic wavefunctions
(projected Fermi liquid) with additional nearest neighbor
correlations input through a Jastrow-type configurational
weighting factor.  The strength of the input correlations
serves as a variational parameter.

Sec.~\ref{sec:PsiJG} studies the properties of these wavefunctions.
We find that up to moderate input correlations, the wavefunction
indeed describes a renormalized Fermi liquid, consistent with
the initial motivation.
We also realize that for strong input correlation and over the 
doping range $0.27 < x < 0.5$ and $0.5 < x < 0.73$, 
our Jastrow-Gutzwiller wavefunction has a $\rt3rt3$ charge order, 
which is inherited from the charge distribution properties of the 
classical Jastrow weight on the lattice.  
Since such a state is beyond our initial motivation, 
we examine its properties and treat it very critically
whenever the variational parameter is driven into this regime.

In Sec.~\ref{sec:RNRMMF} we develop a convenient renormalized meanfield
picture for the energetics in the entire doping range.
We identify the regime of the renormalized Fermi liquid state,
and also the regime where the strong repulsion drives the optimal
Jastrow-Gutzwiller wavefunction into the $\rt3rt3$ charge order
state.

In Sec.~\ref{sec:VMC} we confirm the renormalized meanfield picture
with numerically accurate evaluations with the trial wavefunctions.
We also perform a more detailed study of the possible
$\rt3rt3$ charge order by comparing with the more conventional
CDW states.  We find that our Jastrow-Gutzwiller wavefunction
in the $\rt3rt3$ regime is rather good energetically
and suggest some ways for improving the energetics further.

In Sec.~\ref{sec:RVB} we add the $J$ term and consider the
issue of RVB superconductivity at low dopings.
This is done with the help of the renormalized
meanfield picture.  Without the Jastrow renormalizations,
the RVB superconductivity would not survive to the experimentally
observed $x=0.35$.  We find that the bandwidth suppression
due to charge frustration may indeed resurrect the
superconductivity near $x=1/3$ where such renormalizations
are strongest, particularly if we allow the coexisting
charge order.
We speculate that this may be relevant to explain the narrow 
doping range in which the superconductivity has been found.\cite{Schaak}

Finally, in Sec.~\ref{sec:EXP} we conclude with some simple
predictions for the experiments from the developed charge
frustration picture.  Most notably, transport properties
such as thermopower and Hall effect of the Fermi liquid with low 
degeneracy temperature resemble those of the 
Na$_{0.7}$CoO$_2$ system; these properties look rather unusual
from the perspective of conventional metals.

\vskip 2mm
Before proceeding, we remark about the possibility of charge
order\cite{Ray,Baskaran2} in \nacoo.
The experimental situation is not settled
on this issue.\cite{Gav,Lem,Bru}  We favor the picture
where there is no charge order, but only strong local
correlations.  Charge ordering transition should exhibit
itself in an abrupt change in transport properties,
which has not been observed.

In the present paper, we do spend a lot of time discussing
the particular $\rt3rt3$ order, since it inevitably arises
in our systematic treatment of the concrete model.
We should warn the reader that the details are likely
strongly model-dependent.
Since we do not know the precise microscopic model,
the presented analysis of the charge order should be viewed only 
as an initial sketch of what might happen.
The reported work is done with the nearest neighbor
repulsion (and nearest neighbor Jastrow correlation) only.
Including further neighbor correlations would frustrate
the $\rt3rt3$ charge order and extend the renormalized Fermi
liquid regime, but might also lead to more complicated charge
orders.  We have not pursued such studies systematically,
concentrating on the nearest neighbor case only.

\section{$tV$ model.  Jastrow-Gutzwiller $\Psi$}
\label{sec:tVmodel}
For concreteness, we consider the following single band Hubbard
model with additional nearest neighbor repulsion\cite{Baskaran2}
on a triangular lattice
\begin{equation}
\hat H_{tV} = P_G \sum_{\la ij \ra}
-(t c_{i\sigma}^\dagger c_{j\sigma} + \Hc) P_G
+ V \sum_{\la ij \ra} n_i n_j ~.
\label{HtV}
\end{equation}
Large onsite repulsion is taken into account using the 
Gutzwiller projector $P_G$ to project out double-occupation of 
sites---this gives the ``$t$ part'' as in the familiar $tJ$ Hamiltonian.
We focus primarily on the effect of adding strong
nearest neighbor charge repulsion $V$, and refer to
Eq.~(\ref{HtV}) as $tV$ Hamiltonian.
(We will consider the full $tJV$ Hamiltonian with
$J \ll t, V$ later.)
The band is less than half-filled, with the average
fermion density
$\la c_{i\sigma}^\dagger c_{i\sigma} \ra = \rho < 1$.

\begin{figure}
\centerline{\includegraphics[width=\columnwidth]{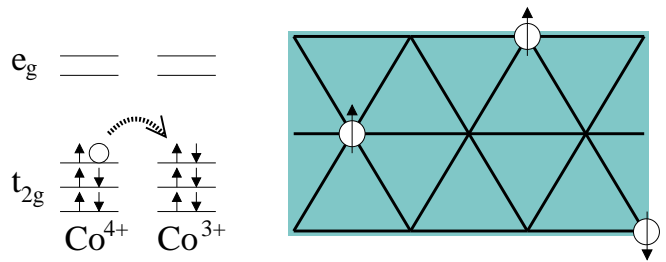}}
\vskip -2mm
\caption{\nacoo: Schematic pictures
(borrowed from Ref.~\onlinecite{WangQ}) explaining the single band
electronic model.
The charge carriers are spin-1/2 charge $q=|e|$ holes of density
$\rho=1-x$.
}
\label{fig:nacoo}
\end{figure}

In the context of the \nacoo system, $c_{i\sigma}^\dagger$
creates a spinful hole and represents the motion of a
Co$^{4+}$($S=1/2$) site, while Co$^{3+}$($S=0$) sites have no holes;
$\rho = 1-x$.  This is shown schematically in Fig.~\ref{fig:nacoo}.
For more details, see also
Refs.~\onlinecite{Baskaran1, KumarShastry, WangLeeLee}.
We take $t>0$, as in Refs.~\onlinecite{Baskaran1, WangLeeLee}.
This is consistent with the photoemission studies,\cite{Has} and
also with the high temperature behavior of the Hall
coefficient\cite{WangRH} (see Sec.~\ref{sec:EXP}).
In this picture the end compound NaCoO$_2$ consists of all 
Co$^{3+}$ with no holes $(\rho = 0, x = 1)$,
whereas the hypothetical end compound Na$_{x=0}$CoO$_2$ consists of a 
Mott insulator with all Co$^{4+}$ sites each carrying 
$S = \frac{1}{2}$ $(\rho = 1, x = 0)$. 
From the latter point of view, Na$_x$CoO$_2$ can be viewed as 
electron doping by a concentration of $x$ electrons into a Mott
insulator.

Here, in order to gauge the effect of the nearest neighbor repulsion,
we perform a trial wavefunction study of this
strongly correlated system.
When $V=0$, a good trial wavefunction is obtained by
Gutzwiller-projecting a simple free fermion state:
\begin{eqnarray*}
|\Psi_G \ra = P_G |\Psi_0 \ra
= \sum_{\{R\}, \{R'\}}
&& \det\left[\phi_i(R_j)\right]
   c_{R_1\uparrow} \dots c_{R_{N/2}\uparrow} \\
\times && \det\left[\phi_i(R_j^\prime) \right]
   c_{R_1^\prime \downarrow} \dots  c_{R_{N/2}^\prime \downarrow} ~,
\end{eqnarray*}
where the sum is over all configurations of spin up and down fermions
with all $R_j$ and $R'_{j'}$ distinct, $\{R\} \cap \{R'\} = 0$.
Away from half-filling, this Gutzwiller wavefunction is a
Fermi liquid state; this can be confirmed, e.g., by measuring
the quasiparticle $Z$ from the step in $\la \hat n_{\bm k}\ra$.
In this state, we have approximately
$\la n_i n_j \ra \approx \rho^2$,
while the fermion kinetic energy can be fairly accurately
estimated from that of the preprojected
free fermions\cite{Vollhardt, FCZhang, Gros}
\begin{equation}
\frac{\la\Psi_G | \hat H_t | \Psi_G\ra}{\la\Psi_G | \Psi_G\ra}
\approx g_t \frac{\la\Psi_0 | \hat H_t | \Psi_0\ra}{\la\Psi_0 | \Psi_0\ra} ~,
\end{equation}
with
\begin{equation}
g_t = \frac{1-\rho}{1-\rho/2} = \frac{2x}{1+x} ~.
\label{gthop0}
\end{equation}
This is commonly referred to as the Gutzwiller approximation, 
or the renormalized mean field theory.
The renormalization factor $g_t$ can be obtained by counting
the number of real space configurations available for hopping
in the projected and preprojected states, and ignoring all
other wavefunction differences.

Turning on the nearest neighbor repulsion $V$, we schematize its effect
on the ground state by introducing an additional Jastrow-type factor
\begin{equation}
\exp \left[- \frac{W}{2} \sum_{\la ij \ra} n_i n_j \right]
\end{equation}
for each real space configuration of fermions in the above
Gutzwiller wavefunction.  $W>0$ effectively suppresses the nearest
neighbor occupation probability, and can be varied to optimize
the trial energy of the $tV$ Hamiltonian.
We will refer to this wavefunction as Jastrow-Gutzwiller (JG)
$\Psi_{JG}$.

It is clear that the effect of $V$ is most severe for 
$x = 1/3$ and $2/3$. 
For $V \gg t$ we expect $W \gg 1$, in which case the 
vacancies (for $x = 1/3$) or the spin carrying holes (for $x = 2/3$) 
would form a $\rt3rt3$ structure to minimize the repulsion.  
Furthermore in this state the particles cannot hop without paying the 
energy $V$.  For intermediate $V$ and away from commensuration
some remnant of this ``jamming'' phenomenon may remain and this is 
what we would like to investigate in this paper.

\section{Properties of $\Psi_{JG}$.  Classical lattice gas system.}
\label{sec:PsiJG}
Before presenting the optimized energetics with $\Psi_{JG}$,
we first discuss its properties.
For nonzero but small $W$, it still describes a Fermi liquid
state with some further renormalizations compared to the
Gutzwiller wavefunction.
However, one has to be careful when $W$ becomes large:
The probability of finding a particular configuration
of charges now has an additional ``classical'' weight
$\exp [- W \sum_{\la ij \ra} n_i n_j]$, and one has to be
wary of the possibility of phase transitions in the
corresponding statistical system.
Factoring out many possible spin assignments for each
charge configuration, we need to consider a classical
system of particles with nearest neighbor repulsion
on a triangular lattice with the classical partition
function
\begin{equation}
Z_{\rm class} = \sum_{\{n_i\}} e^{ -{\EuFrak U}_{\rm class}[n] }
= \sum_{\{n_i\}} e^{ -W\sum_{\la ij \ra} n_i n_j } ~.
\label{Zclass}
\end{equation}
Here $n_i=0,1$, and we work at fixed density $\rho$ as appropriate
for the discussion of our trial wavefunctions with fixed fermion
number.
$W$ plays the role of the inverse temperature in this classical
system, $W=T^{-1}$ (the classical repulsion strength is set to one).

This lattice gas system has been extensively studied
in statistical physics,\cite{MEFisher, Baxter} most notably as a
model for adsorbed monolayers of rare-gas atoms on graphite.
Also, it is equivalent to a triangular lattice Ising
antiferromagnet in an external field; fixed particle density
corresponds to fixed magnetization in the Ising system.

\begin{figure}
\centerline{\includegraphics[width=\columnwidth]{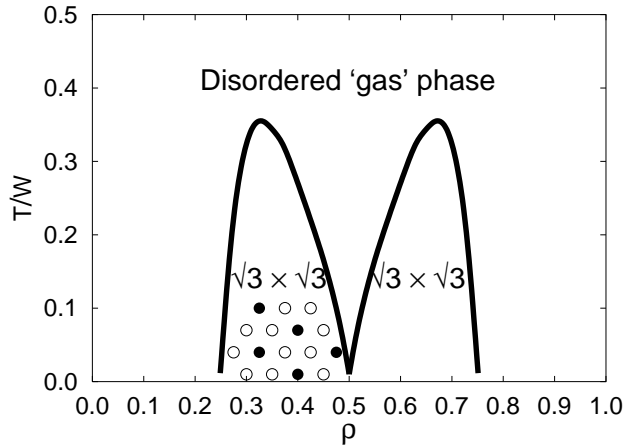}}
\vskip -2mm
\caption{Phase diagram of the classical lattice gas system
Eq.~(\ref{Zclass}).  Note the symmetry relative to $\rho=0.5$ due to
particle-hole symmetry.
}
\label{classical_phased}
\end{figure}

The phase diagram in the $\rho - T$ plane is shown in
Fig.~\ref{classical_phased}.
It is symmetric with respect to $\rho=0.5$ due to particle-hole
symmetry in this classical system, and we discuss the
$\rho<0.5$ part only.

At high temperatures (small $W$) the system is in a disordered
gaseous phase.  For small particle density $\rho < 0.27$
the system remains in the gas phase all the way to zero
temperature.  For densities $0.27 < \rho < 0.5$
the system ``crystallizes'' into a $\rt3rt3$ state at low temperature.
This state is characterized by a preferential particle occupation
of one of the three sublattices of the triangular lattice; the order
is strongest near the commensurate $\rho=1/3,\, 2/3$.
Note that away from these $\rho$ the $\rt3rt3$ phase is more properly
characterized as a density wave state rather than a crystal.
In particular, a fraction of particles remains relatively
mobile having no activation energy for their motion.
The $W=\infty$ model is equivalent to the Baxter's hard hexagon
model and is exactly solvable.\cite{Baxter}

It is useful to have the following caricatures of the charge
motion in the $\sqrt{3}\times\sqrt{3}$ phase, appropriate
at low temperature (large $W$) and near the commensurate
filling $1/3$.  For $\rho>1/3$, we have one sublattice, say $A$,
completely occupied by particles (these particles are almost
localized), while the remaining small density is smeared out
relatively uniformly over the honeycomb lattice formed by the
$B$ and $C$ sublattices.
For $\rho<1/3$, we picture the $B$ and $C$ sites as completely
empty (with almost no density fluctuation),
while all particles are spread over the $A$ sublattice.
The charge motion is achieved by hops from occupied $A$ sites to
neighboring empty $A$ sites via $B$ or $C$ sites;
the most effective such hops involve at least two neighboring empty
$A$ sites in order to avoid the repulsion energy cost for the
intermediate step (see Fig.~\ref{dicelat} in Sec.~\ref{sec:VMC}).

\vskip 2mm
Returning to our trial wavefunction, we expect it to roughly
inherit the charge distribution properties of the lattice gas
system Eq.~(\ref{Zclass}).  Thus, for $W$ such that
the classical system is in its disordered phase,
the wavefunction realizes a Fermi liquid state.
On the other hand, when the classical system is in the
$\sqrt{3} \times \sqrt{3}$ phase, we have checked by 
variational Monte Carlo studies that the resulting wavefunction
has a charge density wave (CDW) order but retains some liquid
properties.  (The transition point $W_c$ is slightly
different for the wavefunction and the classical system.)
Note that the above Jastrow weight realizes a soft projection
satisfying the nearest neighbor repulsion $V$.
It follows that a hard such projection ($W \to \infty$)
leads to the $\rt3rt3$ density wave for
$0.27 < \rho < 0.5$ and $0.5 < \rho < 0.73$.

Before proceeding further, we want to emphasize again that
we set out to study Fermi liquid renormalizations induced by
nearest neighbor repulsion.
Fermi liquid state is achieved in the parameter regime
where the classical system remains disordered; in this
case, our treatment is consistent.

On the other hand, when the nearest neighbor repulsion
is strong, it drives the optimal variational parameter $W$
of $\Psi_{JG}$ into the regime with the CDW order.
In this case, we have to be very cautious in interpreting the
``transition'' and the resulting state, since our original
assumptions about the properties of the wavefunction
no longer hold.  We may still interpret this as a sign
of an instability towards a different state (most likely
with charge order), but the JG wavefunction in this
regime should be treated very critically, particularly since it
has somewhat unusual charge distribution properties.
Thus, one should at least examine other more conventional
trial states with different orders.
This is done in Sec.~\ref{sec:VMC}.

\section{Energetics with $\Psi_{JG}$:  Renormalized meanfield picture.}
\label{sec:RNRMMF}
We now proceed to the actual energetics with the Jastrow-Gutzwiller
trial wavefunctions.
It is possible to perform essentially exact evaluations of the
expectation values with such fermionic wavefunctions using a
well established and documented Variational Monte Carlo (VMC)
procedure.\cite{Ceperley, FCZhang, Gros}
Such detailed studies of the $tV$ energetics
at experimentally relevant $x=0.7$ and $0.35$ are reported in the
next section.

It is also possible to study more complicated Hamiltonians.
However, VMC evaluations are computationally rather costly.
Furthermore, they become inconclusive when the energy differences
become very small.  This is particularly the case when we attempt
to study the physics at energy scales below the dominant $t$ and $V$,
e.g., if we want to resolve the spin sector,
or study pairing instabilities due to the $J$ term.

Useful and fairly accurate guidance is obtained through
the following ``renormalized meanfield''
procedure,\cite{Vollhardt, FCZhang, Gros}
which is much simpler computationally and is also more amenable
to interpretation and extrapolation in the regime where
VMC results become inconclusive.
Generalization of the configuration counting arguments mentioned
earlier leads to the following estimate of the hopping energy
renormalization in the JG wavefunction relative to the unprojected
free fermion wavefunction:
\begin{equation}
\frac{\la\Psi_{JG} | c_{i\uparrow}^\dagger c_{j\uparrow} | \Psi_{JG}\ra}
     {\la\Psi_{JG} | \Psi_{JG}\ra}
= g_t[i,j]
\frac{\la\Psi_0 | c_{i\uparrow}^\dagger c_{j\uparrow} | \Psi_0\ra}
     {\la\Psi_0 | \Psi_0\ra}
\end{equation}
with
\begin{widetext}
\begin{equation}
g_t[i,j] = \frac{1}{\rho (1-\rho/2)}
\left\la\!\!\!\left\la
\delta(n_i-0) \delta(n_j-1)
\exp\left[-\frac{1}{2}
           \left\lgroup {\EuFrak U}_{\rm class}[n_i\!=\!1, n_j\!=\!0]
                  -{\EuFrak U}_{\rm class}[n_i\!=\!0, n_j\!=\!1]
                  \right\rgroup
           \right]
\right\ra\!\!\!\right\ra ~.
\label{gthopW}
\end{equation}
\end{widetext}
Here, $\lala \dots \rara$ denotes averaging in the
classical lattice gas system with the weight
$\sim \exp(-{\EuFrak U}_{\rm class}[n])$ discussed earlier.
When obtaining this expression, similar to the original Gutzwiller
approximation Eq.~(\ref{gthop0}), we again ignored the details of
the fermionic determinant weighting of configurations,
but kept the Jastrow weighting.  Only configurations with the
occupied $j$ site and unoccupied $i$ site contribute, and the
specific ``transition weight'' comes from the corresponding
Jastrow weighting of the configurations before and after the hop.
Note that
${\EuFrak U}_{\rm class}[n_i\!=\!1, n_j\!=\!0]
  -{\EuFrak U}_{\rm class}[n_i\!=\!0, n_j\!=\!1]$
is a local energy term involving only the affected sites $i,j$, and
their immediate neighbors.

Similarly, we can approximate the nearest neighbor repulsion energy
by
\begin{equation}
\frac{\la\Psi_{JG} | \hat n_i \hat n_j | \Psi_{JG}\ra}
     {\la\Psi_{JG} | \Psi_{JG}\ra}
\approx
\lala n_i n_j \rara ~.
\label{Enn}
\end{equation}
The required classical expectation values are readily evaluated
via a Monte Carlo study of the lattice gas system.
As we will see in the next section, such renormalized
meanfield procedure indeed gives fairly accurate estimates of the
expectation values in the Jastrow-Gutzwiller wavefunction, both
in the metallic and the density wave regimes.

\begin{figure}
\centerline{\includegraphics[width=4in]{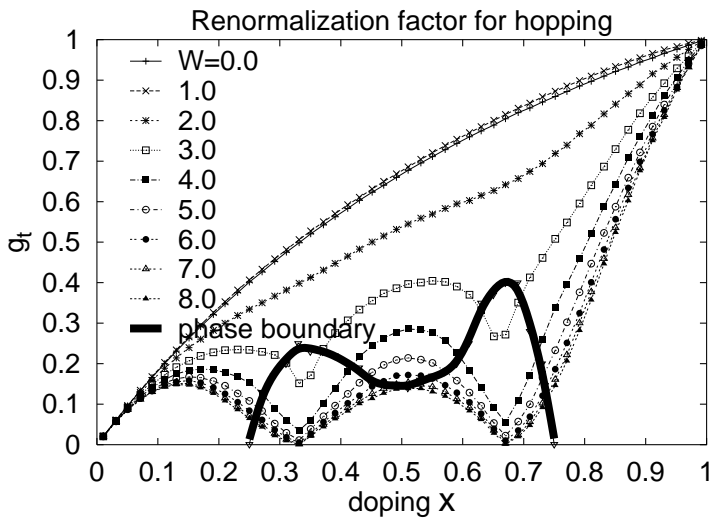}}
\vskip -2mm
\caption{Jastrow-Gutzwiller renormalization factor for hopping,
Eq.~(\ref{gthopW}), as a function of doping for a number of
fixed $W$.  Evaluations are done via classical Monte Carlo
study of the lattice gas system.
The dark thick line delineates the phases of $\Psi_{JG}$.
The $\rt3rt3$ phase lies below the thick line and in this region
the exhibited $g_t$ is averaged over all bonds.
}
\label{fig:gthop}
\end{figure}

\begin{figure}
\centerline{\includegraphics[width=\columnwidth]{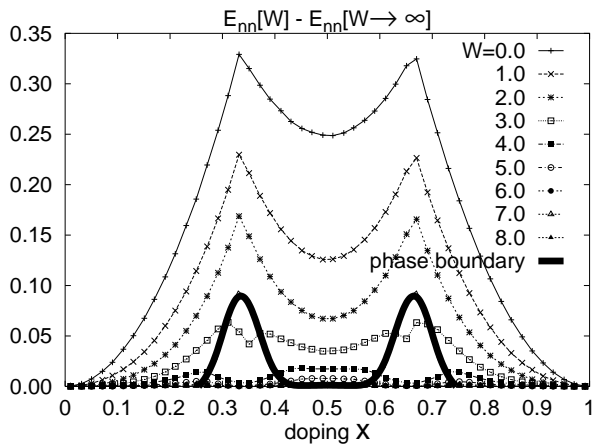}}
\vskip -2mm
\caption{Nearest-neighbor repulsion energy per site,
referenced to the $W\to\infty$ value.
The $\rt3rt3$ charge ordered phases of $\Psi_{JG}$ lie below the 
thick line.
}
\label{fig:Enn}
\end{figure}

We can now develop an overall picture for all fermion densities.
Particular cuts through the results are shown
in Figs.~\ref{fig:gthop} and \ref{fig:Enn}.
Figure~\ref{fig:gthop} shows the hopping renormalization factor
$g_t$ as a function of $x$ for a number of fixed $W$.
(Complimentary cuts through the data for the specific fixed
$x=0.70$ and $x=0.35$ can be also found in the next section.)
Figure~\ref{fig:gthop} is the core of the present paper.

The $W\!=\!0$ curve gives precisely the original Gutzwiller
approximation Eq.~(\ref{gthop0}) for the no-double-occupancy constraint.
This sets a useful reference for gauging the additional effect
of the nearest neighbor repulsion.
The curve with the largest $W=8$, on the other hand, essentially
realizes a complete projection that satisfies the
nearest neighbor repulsion; this is the maximal renormalization
that can be achieved with such nearest neighbor correlations.
The phase boundary of the classical lattice gas
(cf. Fig.~\ref{classical_phased}) is sketched by a thick dark
line:  All points above the line are in the disordered phase
(Fermi liquid wavefunctions),
while points below the line are in the $\rt3rt3$ density wave phase.

Figure~\ref{fig:Enn} shows a similar plot for the repulsion energy
$E_{nn} = \sum_{\la ij \ra} \lala n_i n_j \rara$ per site
(cf. Eq.~\ref{Enn}), which we reference to the minimal possible 
repulsion energy at a given density:
$E_{nn, W\to\infty}=0$ for $\rho \in [0,\, 1/3]$,
$3\rho-1$ for $\rho \in [1/3,\, 2/3]$, and
$6\rho-3$ for $\rho \in [2/3,\, 1]$.
When plotted in this way, the result is symmetric with respect to
$\rho=0.5$ due to classical particle-hole symmetry.
Again, the classical phase boundary is sketched with a
thick dark line.  Observe that the curves with $W>5$
give almost complete ``minimum-nearest-neighbor'' projection.

\begin{figure}
\centerline{\includegraphics[width=\columnwidth]{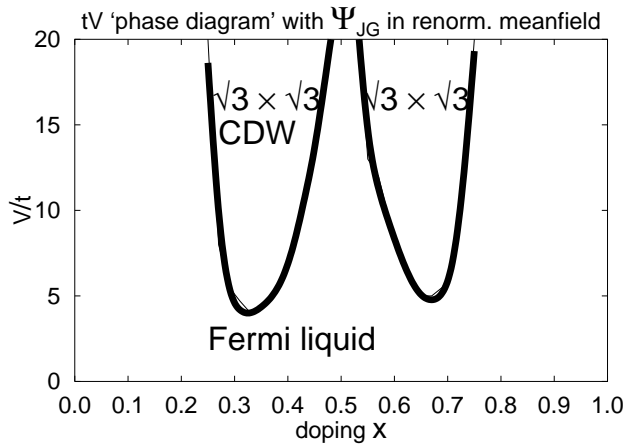}}
\vskip -2mm
\caption{``Phase diagram'' obtained by optimizing the $tV$
Hamiltonian over the JG wavefunctions.
The different ``phases'' correspond to the difference
in the physical properties of $\Psi_{JG}$ as a function of $W$.
Calculations are performed in the renormalized meanfield
approximation using the data of
Figs.~\ref{fig:gthop}, \ref{fig:Enn}, and the free fermion
$\la \hat H_t \ra_0(x)$.
The phase diagram is not expected to be symmetric relative to
$V$ --- cf.~Fig.~\ref{fig:gthop}.  The observed rough symmetry
is due to compensating tendencies in $g_t(x)$ and
$\la \hat H_t \ra_0(x)$ that make the actual kinetic energy
``more symmetric'' relative to $x=0.5$.
}
\label{fig:rnrmJGphased}
\end{figure}

With this data, and also using the free-fermion $\la \hat H_t \ra_0(x)$
(not shown), we can optimize the full $tV$ Hamiltonian in this
renormalized meanfield procedure for $\Psi_{JG}$.
The resulting ``phase diagram'' can be seen in
Fig.~\ref{fig:rnrmJGphased}:  For each doping
$x \in [0.27, 0.73]$ we show the ``critical'' $V/t$ that drives the
optimal $W$ into the regime with the $\rt3rt3$ order.

We emphasize that this ``phase diagram'' is for the optimized
Jastrow-Gutzwiller wavefunction only; in particular, the exhibited
``phase transition'' corresponds to the transition in the properties
of $\Psi_{JG}$ as a function of parameter $W$.
It need not correspond to the actual phase diagram of the
$tV$ Hamiltonian.
It is exhibited here primarily to delineate the regimes
where the JG renormalized Fermi liquid can adequately describe the
$tV$ model energetics, and also where such description is no longer
possible.  In the latter case, one should seriously examine other
physical states paying particular attention to charge order.
Whether the JG wavefunction in its
$\rt3rt3$ phase can adequately describe the possible charge
ordering in the system is a separate question that
requires a detailed study.  We will discuss this more specifically
in the next section.  Here we only note that it is
rather fortuitous that our trial wavefunction with a single
variational parameter exhibits two phases, and the initially
``unexpected'' charge-ordered state should be treated with great
caution.

We now return to the main question of this work---the bandwidth
suppression due to nearest neighbor repulsion.
Again, consider Fig.~\ref{fig:gthop}.
A conservative approach is to insist that we consider
Fermi liquid wavefunctions only.  In this case, we should
disregard the data points that end up in the charge
ordered phase.  We still see that there can be significant
renormalizations by a factor of three to five relative to the
bare hopping amplitude even remaining in the Fermi liquid state.
For a fixed $W$, these renormalizations are strongest near the
commensurate $1/3$ and $2/3$ fillings, and weakest near $x=1/2$.
Also, as can be implied from the ``phase diagram'',
the effect of the nearest neighbor repulsion $V$
is strongest near $x=1/3,\, 2/3$.

On the other hand, if we are to take the Jastrow-Gutzwiller
$\rt3rt3$ density wave regime seriously, there can be even
stronger renormalizations of the hopping energy, particularly
near the commensurate densities.
As we will suggest in the following more specific discussion of
the CDW regime, the entire picture provided by Fig.~\ref{fig:gthop} 
including the data under the phase boundary is indeed useful, 
but may require some less important adjustments.
This is because in the $\rt3rt3$ regime the charge order is such
that there remain mobile (even if strongly constrained) carriers;
there is no charge gap since there is no nesting for
the considered dopings.
The Jastrow-Gutzwiller wavefunction and the above renormalized
meanfield treatment also capture this, and give a first useful
guess on the effect of charge order on the fermion kinetic energy.

\section{Energetics with $\Psi_{JG}$:  VMC study.  Possible CDW.}
\label{sec:VMC}
We now consider the $tV$ model energetics in more detail
for specific $x=0.70$ and $x=0.35$.  These values are
relevant for the unhydrated and hydrated Na$_x$CoO$_2$.
The evaluations with the trial wavefunctions are done essentially
exactly using VMC.\cite{Gros, Ceperley}
This more concrete setting will allow us to discuss some robust
features that emerge from our study vs the specifics of the
particular Hamiltonian used to model charge frustration.
Since an accurate treatment of the CDW states may depend on 
specific details, the present discussion is only intended to give a 
flavor of the possibilities that should be considered.

\subsection{Doping $x=0.70$}
\begin{figure}
\centerline{\includegraphics[width=\columnwidth]{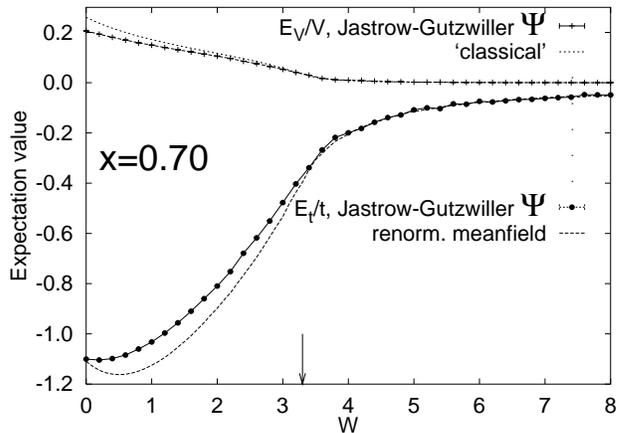}}
\vskip -2mm
\caption{Expectation values of the hopping and nearest neighbor
repulsion energies in the Jastrow-Gutzwiller wavefunction at
doping $x=0.70$ (fermion density $\rho=0.3$).
We also show the corresponding renormalized meanfield values.
Vertical arrow near $W_c \approx 3.3$ indicates
the transition point in the classical lattice gas system.
}
\label{BTWscan_xdopn70}
\end{figure}

Figure~\ref{BTWscan_xdopn70} shows expectation values of the
two parts of the $tV$ Hamiltonian in the Jastrow-Gutzwiller
wavefunction for varying $W$ evaluated using VMC.
It also shows the renormalized meanfield approximation
to these expectation values.  As mentioned in the previous
section, this approximation is indeed fairly accurate
and can be taken seriously.
Since we will be comparing several trial states, we will
use only VMC results in this section.

At this particle density we have $W_c \approx 3.3$ in the
corresponding lattice gas system.
Notice that near $W_c$ the repulsion energy drops quickly and
essentially all the way to zero, similar to the transition in the
classical system.  This is because it is possible to
completely satisfy the repulsion energy by arranging charges so
that there are no nearest neighbors.
Also, such arrangements still allow some fermion hopping, so there
remains nonzero kinetic energy gain even for very large $W$.

\begin{figure}
\centerline{\includegraphics[width=\columnwidth]{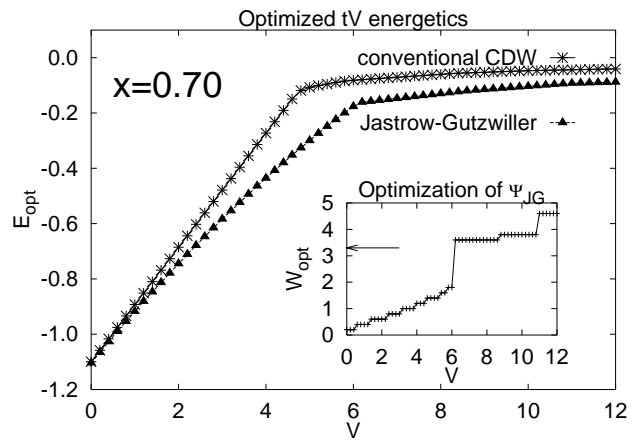}}
\vskip -2mm
\caption{Optimization of the $tV$ Hamiltonian over Jastrow-Gutzwiller
and conventional CDW trial wavefunctions.  Inset shows the optimal
$W_{\rm opt}$ for the Jastrow-Gutzwiller wavefunction;
arrow indicates the critical $W_c \approx 3.3$ in the corresponding
classical lattice gas system.
}
\label{tVopt_xdopn70}
\end{figure}

Using the above data, we can optimize the total energy for different
values of $V/t$.  The result is indicated in Fig.~\ref{tVopt_xdopn70}.
For $V/t \lesssim 6.0$, the optimal $W_{\rm opt}$ remains $\lesssim 2.0$
and the wavefunction is metallic with relatively weak renormalizations.
For larger $V/t$, the optimal $W_{\rm opt}$ jumps to the $\rt3rt3$
ordered side, and $\Psi_{JG}$ has the corresponding density wave order.
Note that the optimal $W_{\rm opt}$ remains fairly close to the
critical value.  In this way, while the repulsion energy
is almost completely satisfied, the system still gains from
some of the original kinetic energy.

We now discuss the regime of large $V$, where the Jastrow-Gutzwiller
energetics suggests charge ordering.
First, it is instructive to compare this with the energetics in a
more conventional CDW trial state.  Such a state is obtained,
for example, by considering a CDW meanfield Hamiltonian
\begin{equation}
H_{\rm CDW}^{\rm mf} \!=\!
-\!\sum_{\la ij \ra} (t_{ij} c_{i\sigma}^\dagger c_{j\sigma} + \Hc)
-\!\sum_i 2 \Delta_{\bm Q} \cos({\bm Q}\cdot {\bm r}_i) \hat n_i ~.
\label{HmfCDW}
\end{equation}
$\Delta_{\bm Q} \equiv \Delta_{\rm CDW}({\bm Q})$ is a CDW order
parameter at the ordering wavevector ${\bm Q}$.  Here,
$\Delta_{\rm CDW}$ serves as a variational parameter for the trial
wavefunction.  In the $\sqrt{3} \times \sqrt{3}$ phase,
the trial $H_{\rm CDW}$ has onsite potential $-2\Delta$ on the
preferred $A$ sublattice and $+\Delta$ on the $B$ and $C$ sublattices
of the triangular lattice.  The trial wavefunction is obtained by
Gutzwiller projection of the meanfield ground state.

The optimized $tV$ energetics for such
more conventional CDW wavefunction is also shown in
Fig.~\ref{tVopt_xdopn70}.
For $V/t \lesssim 4.5$ the optimal wavefunction has
$\Delta_{\rm CDW} \approx 0$, but develops strong CDW order for
larger $V/t$.  In this conventional CDW state at this filling,
we have a coexistence of the charge order and Fermi liquid.

From Fig.~\ref{tVopt_xdopn70}, we see that the JG
wavefunction performs significantly better than the conventional
CDW wavefunction.  This is simple for the metallic side,
since the JG wavefunction has an additional variational parameter
to optimize local correlations compared to the plain metallic state
with $\Delta_{\rm CDW}=0$.
On the other hand, on the charge-ordered side the Jastrow-Gutzwiller
wavefunction performs better almost entirely due to better kinetic
energy.  As discussed earlier, the JG state retains some of the
metallic kinetic energy even in the large $W$ limit.
At the same time, the conventional CDW wavefunction localizes
the fermions to the $A$ sublattice very strongly and loses
essentially all kinetic energy: in the limit of large $V$,
the optimal $\Delta_{\rm CDW} \sim V$ and the optimal total energy
is $\sim -t^2/V$.  Even though the lowest band remains only
partially filled, its bandwidth goes to zero in the limit of
large $\Delta_{\rm CDW}$.

Thus, we conclude that the Jastrow-Gutzwiller wavefunction
with the $\rt3rt3$ order performs fairly well for large $V$.
However, this is by no means the end of the story even for the $tV$
model.
The most serious reservation here is that we have not explored
other competing states in the system for large $V$.  We will
not address this.
We still hope that our approach captures the relevant local
energetics in the system.

\begin{figure}
\centerline{\includegraphics[width=3.0in]{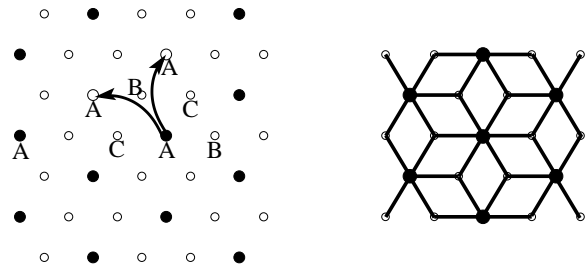}}
\vskip -2mm
\caption{a) Schematics of the $\rt3rt3$ charge order for $\rho<1/3$.
Charges occupy the $A$ sublattice and spend very little time
on the $B$ and $C$ sublattices.  The remaining empty $A$ sites
can be utilized for charge hopping.  There is an intermediate
repulsion energy cost of $V$ to move an isolated such site,
but no such cost for two neighboring empty $A$-sites as shown in
the figure.
b) Dice lattice hopping ansatz motivated by the observation
that hops $B$-$C$ are rarely used.
}
\label{dicelat}
\end{figure}

\begin{figure}
\centerline{\includegraphics[width=\columnwidth]{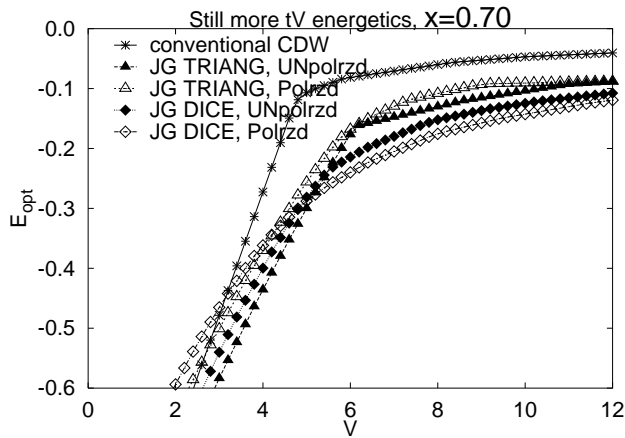}}
\vskip -2mm
\caption{This is a blow-up of Fig.~\ref{tVopt_xdopn70} focusing
on the $\rt3rt3$ regime and showing additional Jastrow-Gutzwiller type
trial wavefunctions for the $tV$ Hamiltonian.
Besides the unpolarized triangular lattice hopping ansatz,
we also show the optimized energetics for the corresponding
fully polarized state, and also for the dice lattice hopping
ansatz.
}
\label{more_tVopt_xdopn70}
\end{figure}

In the present context, we can explore the energetics of the
$\rt3rt3$ ordering more systematically.  As discussed,
the complete minimum-nearest-neighbor projection leads to the
$\rt3rt3$ order.  For $\rho<1/3$, we essentially have charges
living on the $A$ sublattice only and moving primarily via
$A-B-A$ or $A-C-A$ routes, while the bonds $B-C$ are rarely used
(see Fig.~\ref{dicelat}).
In the above, we were projecting the uniform free fermion triangular
lattice hopping ground state, while it is clear that in the resulting
charge ordered state the hops $B-C$ are poorly utilized, and more
generally the kinetic energy---the driving force for uniformity---is
less important.
In the $\rt3rt3$ regime, it then seems more appropriate to project a
hopping state with strong $A-B$ and $A-C$ hopping amplitudes
and weak $B-C$ hops.  The limiting case is the dice lattice
hopping shown in Fig.~\ref{dicelat}b; the six-coordinated
sites are the $A$ sites, while the three-coordinated sites are
the $B$ and $C$ sites.  The dice hopping state by construction
has $\rt3rt3$ order.  It is easy to verify that in the lowest
band one half of the fermion density is located on the
$A$ sublattice.  Clearly, this has better repulsion energy than
the triangular hopping state.  To further optimize the nearest neighbor
correlations, we can introduce a Jastrow weighting as for the
uniform hopping state (note that in the dice case the classical lattice
gas transition is no longer relevant since the charge system
has the $\rt3rt3$ order from the outset).  The optimized $tV$
energetics is shown in Fig.~\ref{more_tVopt_xdopn70},
and we indeed find that the dice hopping ansatz is somewhat
better than the uniform state.

Finally, we should point out that we have completely ignored the spin
physics by considering only unpolarized wavefunctions.
It should be clear that since the bandwidth becomes so narrow,
there will be significant degeneracy---on the $tV$ energy scale---in
the spin sector.  This degeneracy will be resolved in some way or other
at lower energy scale, and the details will depend largely on the
specifics of the microscopic Hamiltonian.
%
%
As an example, trying out spin-polarized Jastrow-Gutzwiller wavefunctions
in the $tV$ Hamiltonian, we find that in the charge-ordered regime
the fully polarized wavefunction performs only slightly worse
than the unpolarized one.  For the dice hopping ansatz, on the other
hand, the spin-polarized wavefunction performs better than the
unpolarized one.
One can get some feeling of the slight differences by examining
Fig.~\ref{more_tVopt_xdopn70}.
Such itinerant ferromagnet tendencies become even more pronounced at
lower fermion density (higher $x$).

\vskip 2mm
After the presented detail, it should be clear that the
energetics can be rather subtle and model dependent,
particularly in the charge order regime.
We now want to separate out which features are more
robust than the above specifics.  This is important
since at present we do not have a good knowledge of the
microscopic Hamiltonian in the \nacoo system.

First of all, we conclude that there can be significant
renormalizations in the metallic wavefunction.
The hopping can be effectively suppressed roughly by a factor of
three (see Fig.~\ref{BTWscan_xdopn70}),
with the wavefunction retaining its Fermi liquid character.
The achievable Fermi liquid renormalizations may be even larger if we
include further neighbor repulsion, since this will frustrate the
$\sqrt{3}\times\sqrt{3}$ charge order and give more parameter
space to the liquid state with uniform charge distribution.
As long as the system remains uniform, this is not sensitive to the
microscopics.  (At this local ``high energy'' level of analysis we
completely disregard the low energy instabilities of the resulting
Fermi liquid state.)

Our second observation is about the nature of possible charge
orders in such strongly frustrated system.
Our JG wavefunctions offer an interesting possibility of essentially
satisfying the nearest neighbor Coulomb repulsion while retaining
some kinetic energy gain and metallicity.
Projecting the triangular or dice lattice hopping is merely a
detail of how the quantum tunneling is put into the wavefunction,
but the overall picture of the resulting state is the same.
Whether such state is energetically favorable compared with
other competing states requires a more detailed study.

Finally, we expect that the spin dynamics is highly degenerate
in such charge frustrated systems, and its ultimate fate is
resolved only at much lower energy scales.

\subsection{Doping x=0.35}
We now summarize similar $tV$ study at $x=0.35$.
This is of interest for the hydrated compound
Na$_{0.35}$CoO$_2 \cdot 1.3$H$_2$O that was found to exhibit
superconductivity.

\begin{figure}
\centerline{\includegraphics[width=\columnwidth]{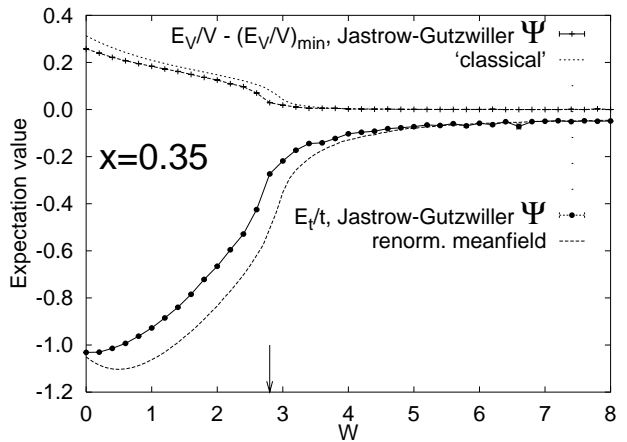}}
\vskip -2mm
\caption{This is similar to Fig.~\ref{BTWscan_xdopn70}, but for
doping $x=0.35$ (fermion density $\rho=0.65$).  The repulsion energy
is referenced to the minimal repulsion energy.
At this density, $W_c \approx 2.8$.
}
\label{BTWscan_xdopn35}
\end{figure}

Figure~\ref{BTWscan_xdopn35} shows the expectation values of the
kinetic and nearest neighbor repulsion energies in the $\Psi_{JG}$
evaluated using VMC.  The repulsion energy is referenced to the
minimal repulsion energy at this density
[$E_{nn, min}=V(3\rho-1)$ per site; cf.~Fig.~\ref{fig:Enn}].
The renormalized meanfield approximation is also shown and is
fairly accurate.

\begin{figure}
\centerline{\includegraphics[width=\columnwidth]{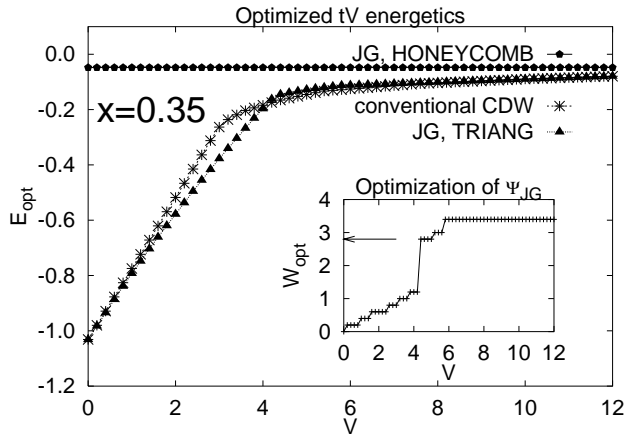}}
\vskip -2mm
\caption{Optimization of the $tV$ Hamiltonian over Jastrow-Gutzwiller
and conventional CDW trial wavefunctions at $x=0.35$
(cf.~Fig.~\ref{tVopt_xdopn70}).
We also show the result for Jastrow-Gutzwiller honeycomb hopping
ansatz.
}
\label{tVopt_xdopn35}
\end{figure}

The result of the wavefunction optimization for the $tV$ Hamiltonian
is shown in Fig.~\ref{tVopt_xdopn35}.
For $V/t \lesssim 4.2$, the optimal $W_{\rm opt}$ remains $\lesssim 1.5$;
the wavefunction is metallic with weak renormalizations.
For larger $V/t$, the optimal $\Psi_{JG}$ jumps to the $\rt3rt3$
ordered side; however, the optimal $W_{\rm opt}$ remains fairly close to
$W_c \approx 2.8$, and the system retains significant part of the
original kinetic energy.

Turning to the regime of large $V$, we consider also
the more familiar CDW trial state obtained from the meanfield
Hamiltonian Eq.~(\ref{HmfCDW}).  The optimized energetics
with such conventional CDW wavefunction is also shown
in Fig.~\ref{tVopt_xdopn35}.  The optimal $\Delta_{\rm CDW}$
remains close to zero for $V \lesssim 3$, but becomes significant
and negative for larger $V$.
We now observe that in the regime with the putative $\rt3rt3$
charge order both the JG wavefunction and the meanfield CDW
wavefunction give very close optimal energies.
This is also true for the individual $t$ and $V$ parts,
suggesting that the Jastrow-Gutzwiller and conventional CDW
wavefunctions give in fact essentially the same physical state.

This can be understood by examining the meanfield CDW state.
For $\Delta_{\rm CDW}<0$ the $B$ and $C$ sublattices are
preferentially occupied, while the $A$ sublattice is preferentially
empty.  In this case, the lower two meanfield bands retain much of the
original bandwidth even in the limit of large $\Delta_{\rm CDW}$.
This is because the $B$ and $C$ sites form a connected
honeycomb lattice, and for large $\Delta_{\rm CDW}$
the two bands correspond essentially to hopping on
this lattice.  The physical state is now obtained
by the Gutzwiller projection of this free fermion honeycomb
lattice hopping state.  But this is also roughly the picture of the
Jastrow-Gutzwiller wavefunction in the $\rt3rt3$ regime for this
density.

The above suggests that we also try projecting honeycomb hopping
ansatz, since it better utilizes the $B-C$ hops.
However, for the range of $V/t$ studied here, the uniform
triangular hopping ansatz performs better, primarily since
it manages to retain some of the $A-B$ and $A-C$ hopping energy.
This completes our exploration of the $\rt3rt3$ order.

Finally, we note that at this high fermion density
$\rho=0.65$, unlike the case with $\rho=0.30$,
the spin degeneracy does not occur, and the unpolarized
wavefunctions are always better.

\vskip 2mm
To summarize, the local energetics of the $tV$ model at $x=0.35$
is well captured by either the renormalized Fermi liquid,
or the $\rt3rt3$ charge ordered state, depending on the value of $V/t$.
The $\rt3rt3$ state also has mobile fermions occupying primarily the
honeycomb sublattice (of the original triangular lattice);
however, since the fermion density is close to complete covering
of the honeycomb lattice, the fermion hopping is strongly
suppressed.

Again, the ultimate fate of the Fermi liquid (or the liquid part in
the $\rt3rt3$ regime) is resolved only at lower energies.
In the following section we study the superconducting instability
due to the $J$ term, and whether the RVB superconductivity can be
significantly enhanced by the discussed strong kinetic energy
suppression.

\section{RVB superconductivity: resurrection near $x=1/3$?}
\label{sec:RVB}
We now turn to the issue of RVB superconductivity due to
the antiferromagnetic spin interaction at dopings $0<x<0.4$.
In the context of the triangular lattice $tJ$ model,
this was considered by several
authors.\cite{Baskaran1, KumarShastry, WangLeeLee, Ogata, Honerkamp}
These studies predict $d+id$ superconductivity.
As expected for such RVB scenario, the superconductivity is
strongest (relative to the metallic state) near half filling
$x=0$, where the charge mobility is very low.
Away from half-filling at moderate dopings,
the need to satisfy the kinetic energy of the carriers leads
to strong suppression of the superconductivity.
As pointed out in Ref.~\onlinecite{WangLeeLee} and discussed further
below, the experimentally observed superconductivity at doping
$x=0.35$ represents a significant problem to this scenario:
If one uses the LDA bandwidth to estimate
$| t_{\rm bare} | \approx 50-100$~meV,
and takes the hopping integral sign as in this work,
and makes a reasonable guess $J \sim 10$ to $20$~meV,
the resulting RVB superconductivity is vanishingly weak for
this doping and would not be observed.

As discussed above---cf.~Fig.~\ref{fig:gthop}---charge
frustration can lead to strong suppression of the
effective hopping amplitude $t_{\rm eff}$, even for
larger doping.
Here we study whether this suppression can be strong
enough to resurrect the superconductivity at $x=0.35$.
Figure~\ref{fig:gthop} also suggests that the region near
$x=1/3$ is special in that it allows the strongest such
renormalizations, with or without the charge ordering.
As discussed earlier, this is because the charge system
is most sensitive to the further neighbor interactions
near this commensurate doping.
On the other hand, when the superconductivity is weak,
the transition temperature is exponentially sensitive
to the effective hopping amplitude (see below).
Thus, we may speculate about the possibility of a
small superconducting dome around this special doping
due to charge correlation (possibly, charge ordering).

The physics treatment presented below is very schematic.
We will essentially think only in terms of the
renormalized couplings and ignore the fact that the underlying
``liquid'' state may be charge-ordered.
This is done to get a rough feeling as to whether
the suggested scenario can work at all.
If the superconductivity near $x=1/3$ can reappear only
in the charge-ordered region, the obtained insight is still
useful and may suggest a more careful treatment.
For example, the charge order may be suppressed by longer range 
repulsion.  Another possibility is that a more accurate treatment 
may lead to short range charge order for intermediate coupling
which shares the same kinetic energy suppression.

We first summarize the standard RVB meanfield for the pure
$tJ$ model.  We formulate this meanfield as an approximate variational
procedure.\cite{FCZhang, Gros, Ogata}
This is particularly convenient for the present work,
which also takes the variational wavefunction perspective.

To study the possibility of singlet superconductivity,
we consider ``trial'' Hamiltonian
\begin{eqnarray*}
H_{\text{trial}} &=& \sum_{ij}
\left[-\chi_{ij} c_{i\sigma}^\dagger c_{j\sigma}
      -(\Delta_{ij} c_{i\uparrow}^\dagger c_{j\downarrow}^\dagger + \Hc)
\right] \\
&-& \sum_i \mu \left[c_{i\sigma}^\dagger c_{i\sigma} - (1-x)\right]~,
\end{eqnarray*}
with $\chi_{ji}=\chi_{ij}^*$, $\Delta_{ji}=\Delta_{ij}$.
For each such trial Hamiltonian, we obtain the corresponding
ground state $|\Psi_0 \ra$.  In the meanfield, we ignore the
no-double-occupancy constraint and only require the average
density to be correctly
$\la c_{i\sigma}^\dagger c_{i\sigma} \ra = 1-x$,
which is achieved by tuning the chemical potential $\mu$.
Going beyond the meanfield, the physical wavefunction is
obtained by Gutzwiller projection.

As discussed at length earlier, we can approximate the
expectation value of the $tJ$ Hamiltonian in the physical
wavefunction by proper renormalizations of the meanfield
values:
\begin{widetext}
\begin{equation}
\frac{\la\Psi_G | \hat H_{tJ} | \Psi_G\ra}{\la\Psi_G | \Psi_G\ra}
\!\approx\!
g_t \frac{\la\Psi_0 | \hat H_t | \Psi_0\ra}{\la\Psi_0 | \Psi_0\ra}
+ g_J \frac{\la\Psi_0 | \hat H_J | \Psi_0\ra}{\la\Psi_0 | \Psi_0\ra}
\!=\!
-g_t \sum_{\la ij \ra} t_{ij}
     \la c_{i\sigma}^\dagger c_{j\sigma} \ra + {\rm c.c.}
-g_J \sum_{\la ij \ra} \frac{3 J_{ij}}{8}
     \left[ |\la c_{i\sigma}^\dagger c_{j\sigma} \ra|^2
           +|\la \epsilon_{\sigma\sigma'}
                 c_{i\sigma}^\dagger c_{j\sigma'}^\dagger \ra|^2 \right]
~.
\label{Emf}
\end{equation}
\end{widetext}
The hopping renormalization factor is given by Eq.~(\ref{gthop0}),
while for the Heisenberg exchange we have\cite{FCZhang, Gros}
\begin{equation}
g_J = \frac{4}{(1+x)^2} ~.
\end{equation}
These estimates follow essentially from the no-double-occupancy
configuration constraints, and do not depend on the
details of the preprojected state as long as it is spatially uniform.
Also, they give numerical results that are fairly close to the
actual evaluations with the projected wavefunctions, as discussed
earlier.
The above is precisely the renormalized meanfield formulation
of Refs.~\onlinecite{FCZhang, Gros, Ogata}.
The slave boson meanfield of Ref.~\onlinecite{WangLeeLee} uses instead
$g_t=x$ and $g_J=1$, so their numerical values are somewhat different.

In this formulation, only the ratio $\Delta/\chi$ is meaningful.
A convenient procedure to minimize Eq.~(\ref{Emf}) is to minimize
instead the so called meanfield Hamiltonian
\begin{equation*}
\hat H_{\rm mf} = \sum_{\la ij \ra} \frac{8}{3 g_s J_{ij}}
\left[ |\chi_{ij}-g_t t_{ij}|^2 + |\Delta_{ij}|^2 \right]
+ \hat H_{\rm trial} ~.
\end{equation*}
By standard arguments, the global minimum of the meanfield
Hamiltonian is also the minimum of the trial expectation
value Eq.~(\ref{Emf}).  In this formulation, the optimal
$\chi$ and $\Delta$ each obtain physical scale as set by
$t$ and $J$.  Thus, we can get a rough idea about the
quasiparticle spectrum above the ground state by considering
the meanfield excitation spectrum, which now has proper physical
scale.
In particular, the optimal $\Delta$ gives a physical measure
of the strength of superconductivity.

The self-consistency conditions read
\begin{eqnarray}
\chi_{ij}^* &=& g_t t_{ij}
+ \frac{3 g_s J_{ij}}{8} \la c_{i\sigma}^\dagger c_{j\sigma} \ra ~,
\label{selfcons_chi}
\\
\Delta_{ij}^* &=& \frac{3 g_s J_{ij}}{8}
\la \epsilon_{\sigma\sigma'}
    c_{i\sigma}^\dagger c_{j\sigma'}^\dagger \ra ~.
\label{selfcons_Delta}
\end{eqnarray}

From now on, we specialize to the $d+id$ superconductor ansatz:
\begin{equation}
\Delta_{{\bm e}_1} = \Delta; \;\;
\Delta_{{\bm e}_2} = \Delta e^{i 2\pi/3}; \;\;
\Delta_{{\bm e}_3} = \Delta e^{i 4\pi/3}.
\end{equation}
Here ${\bm e}_1 = \hat {\bm x}$,
${\bm e}_2 = \frac{1}{2}\hat{\bm x} + \frac{\sqrt{3}}{2}\hat{\bm y}$,
and ${\bm e}_3 = {\bm e}_2 - {\bm e}_1$ are the unit triangular
lattice vectors.  There is strong evidence that this state wins the $tJ$
model energetics for the considered dopings, at least on the meanfield
level.\cite{Baskaran1, KumarShastry, WangLeeLee, Ogata, Honerkamp}

We give the results for $t=3 J$ and $t = 5 J$.  
This is somewhat different from the cited $t=5-10 J$ 
values.\cite{WangLeeLee}  
At the moment, there is significant
uncertainty in the precise microscopic model, while the
superconductivity energy scale is exponentially sensitive to the
microscopic values and to numerical constants in the theory.  
The $t=3J$ results make the demonstration of principle more dramatic. 
A similar, but weaker, effect is seen for $t = 5 J$

The optimal $\Delta$ in units of $J$ is shown as a function of
doping in Fig.~\ref{mfDiD0_t3}.
For weak $J$, the optimal $\Delta$ is given by a BCS-like formula
(see Appendix~\ref{app:DeltaBCS} for details)
\begin{equation}
\Delta \sim t_{\rm eff} e^{-c t_{\rm eff}/J_{\rm eff}}
\label{DeltaBCS}
\end{equation}
with some numerical constant $c=c(x)$
and $t_{\rm eff} = g_t t$, $J_{\rm eff} = g_J J$.
The effective mobility of charges increases with doping
$t_{\rm eff} \sim x t$, and this leads to the observed very
quick drop of $\Delta$.

\begin{figure}
\centerline{\includegraphics[width=\columnwidth]{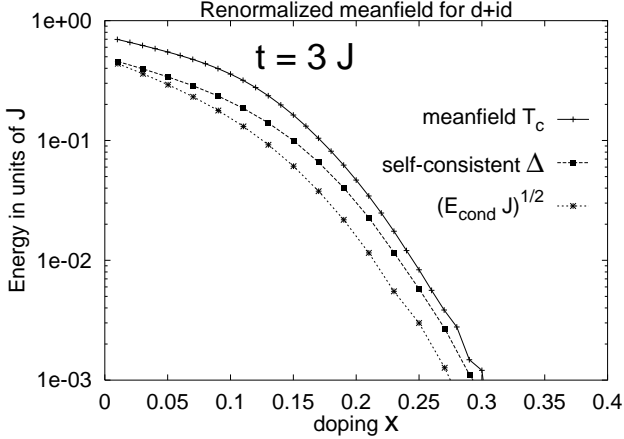}}
\vskip -2mm
\caption{Renormalized meanfield results for the $d+id$ superconducting
state for $t=3J$.  We show the meanfield $T_c$, the optimal $\Delta$,
and the square root of the condensation energy $E_{\rm cond}$.
The energy scale is the bare $J$; note the logarithmic scale for
the energy.
}
\label{mfDiD0_t3}
\end{figure}

Figure~\ref{mfDiD0_t3} also shows two other physical
measures of the strength of superconductivity.
One is the meanfield $T_c$ defined here as the
transition temperature for the finite temperature
optimization of the meanfield Hamiltonian $H_{\rm mf}$.

The other measure is obtained by considering the condensation
energy of the superconducting state.  This is defined as
the energy gain in the optimal superconducting state
relative to the Fermi liquid state ($\Delta=0$).
For small $\Delta$, the condensation energy is expected
to scale as $E_{\rm cond} \sim \Delta^2 / t_{\rm eff}$;
to compare with $\Delta$ in Fig.~\ref{mfDiD0_t3},
we plot instead $(E_{\rm cond} J)^{1/2}$.

From Fig.~\ref{mfDiD0_t3}, these measures all trail each other.
The figure has been somewhat arbitrarily cutoff at $10^{-3} J$.
For $J \sim 20$meV, any $T_c$ or $\Delta$ below this scale would
not be observed in the experiments.
Note the precipitous drop in the strength of superconductivity for
$x \gtrsim 0.20$.  There is simply no hope for it surviving to the
experimentally observed $x=0.35$ in this setting.

We remark here that a direct VMC study must see the condensation
energy to establish the ground state $\Delta$.  Since $E_{\rm cond}$
is extremely small, such studies become impractical.  This is where
the renormalized meanfield procedure becomes very useful.

\begin{figure}
\centerline{\includegraphics[width=\columnwidth]{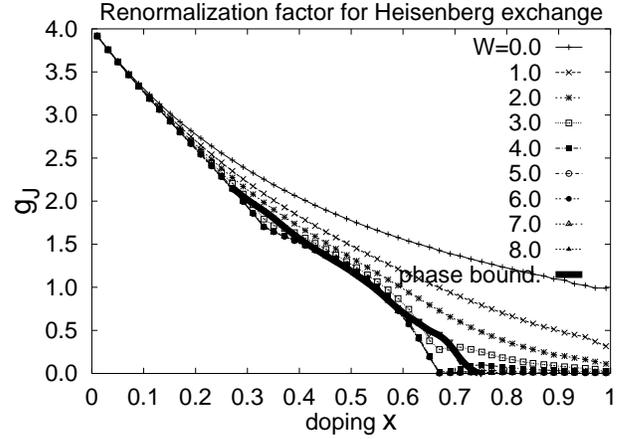}}
\vskip -2mm
\caption{Jastrow-Gutzwiller renormalization factor for the
Heisenberg exchange as a function of doping for a number of
fixed $W$ (cf Figs.~\ref{fig:gthop},\ref{fig:Enn}).
In the $\rt3rt3$ phase (data points below the thick dark line),
the exhibited $g_J$ is averaged over all bonds.
}
\label{fig:gJHeis}
\end{figure}

\vskip 2mm
Let us now return to the $tJV$ model with strong nearest neighbor
repulsion.  We think roughly as follows.
The dominant $t$ and $V$ parts can be satisfied as above
by the appropriate Jastrow weighting of charge configurations
in our trial wavefunctions.  As discussed earlier, the effect of the
Jastrow factor can be conveniently described by the corresponding
renormalizations of the hopping amplitude $g_t$
(Eq.~\ref{gthopW} and Fig.~\ref{fig:gthop}) and the
Heisenberg exchange $g_J$.
The latter is approximated by
\begin{equation}
g_J = \frac{\lala \delta(n_i-1) \delta(n_j-1) \rara}
           {[\rho (1-\rho/2)]^2} ~,
\end{equation}
and is plotted in Fig.~\ref{fig:gJHeis}
(cf.~discussion following Eq.~\ref{gthopW}).
As long as the charge distribution remains uniform,
these renormalizations capture the main effect
of the nearest neighbor correlations built in by the Jastrow
factor.

\begin{figure}
\centerline{\includegraphics[width=\columnwidth]{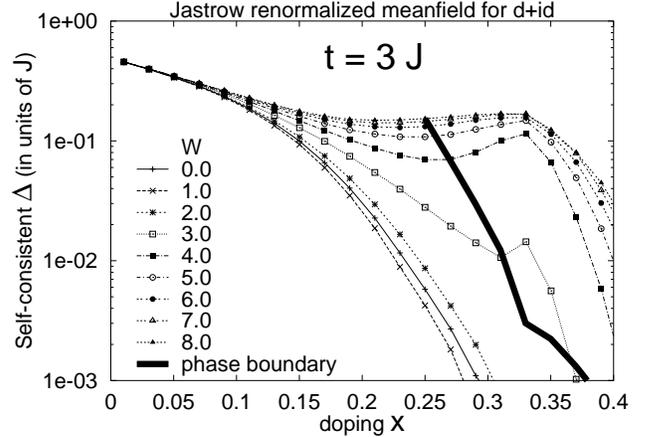}}
\vskip -2mm
\caption{Renormalized meanfield for the Jastrow-weighted
$d+id$ superconducting state for $t=3 J$ (cf.~Fig.~\ref{mfDiD0_t3}).
We show the self-consistent $\Delta$ as the measure of the
superconductivity strength ($T_c$ plots look very similar).
The dark thick line corresponds to the phase boundary of the
Jastrow weight
(cf. Figs.~\ref{classical_phased},\ref{fig:gthop},\ref{fig:gJHeis}) ---
the maximum enhancement of the superconductivity while remaining
in the uniform phase.
}
\label{mfDiDWCH_t3}
\end{figure}

Thus, for each doping level $x$ and the Jastrow
suppression strength $W$, we can estimate the corresponding
$t_{\rm eff}, J_{\rm eff}$, and then the optimal $\Delta$.
The latter is our main measure of the superconductivity strength
and is shown in Fig.~\ref{mfDiDWCH_t3}.
The $W=0$ line is the same as in Fig.~\ref{mfDiD0_t3}, while
the $W=8$ corresponds essentially to the minimum-nearest-neighbor
projection.  Again, the dark thick line corresponds to the phase
boundary of the Jastrow weight.  For $x > 0.27$ all points
above this line have the $\rt3rt3$ charge order.
These are obtained by using the corresponding formal renormalization
factors and the above prescription, even though this violates the
initial motivation coming from a uniform renormalized liquid picture.
As emphasized earlier, the precise energetics in this regime
likely requires a more careful treatment.  However, we expect that even
such simplistic analysis in the $\rt3rt3$ regime gives a reasonable
first guidance on how the role of the kinetic energy can be suppressed
by possible charge ordering in the system.

\begin{figure}
\centerline{\includegraphics[width=\columnwidth]{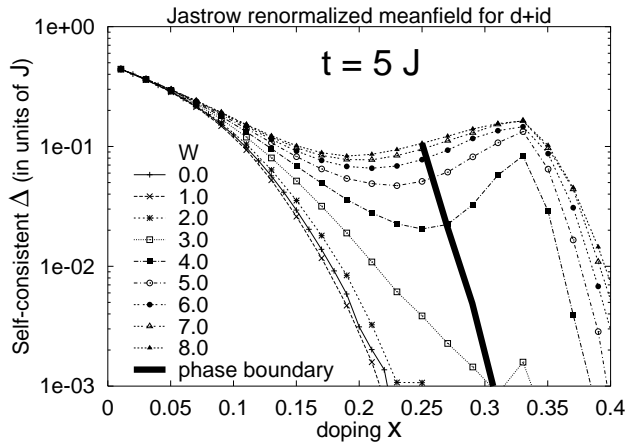}}
\vskip -2mm
\caption{This is the same as Fig.~\ref{mfDiDWCH_t3}, but for $t=5J$.
}
\label{mfDiDWCH_t5}
\end{figure}

Our tentative conclusion from Fig.~\ref{mfDiDWCH_t3} is that
for $t=3J$ the considered nearest neighbor Jastrow renormalizations
that leave the underlying liquid wavefunction in the pure Fermi liquid
state are borderline sufficient to explain the superconductivity
near $x=1/3$.
One should of course judge this critically because of the
exponential sensitivity to the actual value of the ratio
$t_{\rm eff}/J_{\rm eff}$, Eq.~(\ref{DeltaBCS}).
The trend for increasing $t/J$ can be seen by comparing
$t=3J$ Fig.~\ref{mfDiDWCH_t3} and $t=5J$ Fig.~\ref{mfDiDWCH_t5}.

We speculate that in Na$_x$CoO$_2 \cdot 1.3$H$_2$O the
actual situation is qualitatively close to the curve with $W=3$
in Fig~\ref{mfDiDWCH_t3},
which near $x=1/3$ roughly corresponds to the critical strongly
correlated liquid of the nearest neighbor Jastrow weight
(cf. Figs.~\ref{classical_phased},\ref{fig:gthop}).
This curve may lie inside the liquid phase for longer ranged
Jastrow weight.  Another possibility is that only short range 
charge order is developed for intermediate coupling.
One thing should be clear from 
Figs.~\ref{mfDiDWCH_t3},\ref{mfDiDWCH_t5}:
there can indeed be significant enhancement---one to
two orders of magnitude---in the superconductivity scale due to the
kinetic energy suppression in the charge-correlated liquid.
Because the charge system is most responsive near the
commensurate $x=1/3$, this enhancement may be strongest near this
doping, which may explain the experimentally suggested\cite{Schaak}
superconducting ``dome'' around $x=1/3$.
However, we note that even for $W = 3$, $\Delta$ in 
Fig.~\ref{mfDiDWCH_t3} shows only a shallow maximum
near $x = 1/3$.  A possible explanation of the experiment is 
that $x$ significantly less than $1/3$ is not achievable due to 
chemical reasons and superconductivity is simply cut-off.  
Again, the important message we draw from Fig.~\ref{mfDiDWCH_t3} 
is the possibility of pushing $T_c$ up to an observable level near 
$x = 1/3$.

Finally, if we continue the theory into the Jastrow $\rt3rt3$ charge 
order regime,
the $T_c$ enhancement may be even stronger.  This is not surprising,
since the charge mobility is suppressed even further in this case.
Thus, our earlier analysis tells us that for $x>1/3$ we are
essentially doping a nearly half-filled honeycomb lattice.
This picture also suggests some possibilities of treating the
$\rt3rt3$ regime more carefully, similar to our discussion in
Sec.~\ref{sec:VMC}.  For example, for $x > 1/3$,
we can view the fermions as restricted primarily to the honeycomb
lattice.  On the other hand, the $d+id$ state wins the
energetics in the original uniform triangular lattice meanfield 
and needs to be re-examined in the present context. 
The above renormalized meanfield procedure roughly corresponds
to restricting the $d+id$ ansatz onto the honeycomb lattice.
Of course, one should also consider other possible RVB
superconductor states on the honeycomb lattice and decide
which one is optimal energetically.  More generally,
one may want to consider triangular lattice superconducting ansatz
with broken translational symmetry patterned after the $\rt3rt3$
state.  We are not pursuing such studies here, since it is important
to first establish whether the charge ordering occurs at all in the
material.
If this indeed happens, the above rough considerations can give us
some initial idea about the scale of the superconducting instabilities
in such state.

\section{Conclusions.  Connection with experiments.}
\label{sec:EXP}
We conclude by stating some consequences of the discussed
effects of charge frustration.

1) It will clearly be interesting to look for signs of charge order 
near $x=1/3$ and $2/3$ using X-ray or neutron scattering.  
The conductivity is metallic and in the case of
$x = 1/3$ reaches $50 e^2/h$ at low temperatures.\cite{Chou,WangQ}  
This suggests that long-range charge ordering is unlikely, 
but there may be a tendency for short-range ordering.

2) There can be strong suppression of the effective hopping
amplitude due to nearest neighbor repulsion while remaining
in the Fermi liquid state.
The meanfield hopping amplitude $\chi_{ij}=\chi$ is given by 
Eq.~(\ref{selfcons_chi}) and has contributions proportional to 
$g_t t$ and $g_J J$.  Note that in addition to the suppression of 
$g_t$ (Fig.~\ref{fig:gthop}), $g_J$ is also suppressed 
(Fig.~\ref{fig:gJHeis}), especially for $x > 0.5$.

This suppression leads to low fermion degeneracy temperature.
The properties of such Fermi liquid system with
$\epsilon_F \lesssim T$ are rather unusual from the perspective
of the familiar metals with $\epsilon_F \gg T$
(the Fermi energy is measured from the bottom of the band,
and is roughly $\epsilon_F \sim t_{\rm eff}$).

a) In particular, the thermopower is large and saturates to
the value
\begin{equation}
Q = -\frac{\mu}{qT} = \frac{k_B}{q} \ln\frac{2-\rho}{\rho} ~.
\label{Q:highT}
\end{equation}
at large temperature.  Note the ``classical'' scale
$k_B/|e| = 86.2$ $\mu$V/K, which is in fact observed in
\nacoo.\cite{Terasaki, Koshibae, WangQ}
The full temperature dependence
for $x=0.70$ is shown in Fig.~\ref{QRHx70}.
Here and below, we use simple-minded transport theory
summarized in Appendix~\ref{app:transport}.
Form Fig.~\ref{QRHx70}, the thermopower reaches one half of
the maximal value for $T \approx t_{\rm eff}$.

\begin{figure}
\centerline{\includegraphics[width=\columnwidth]{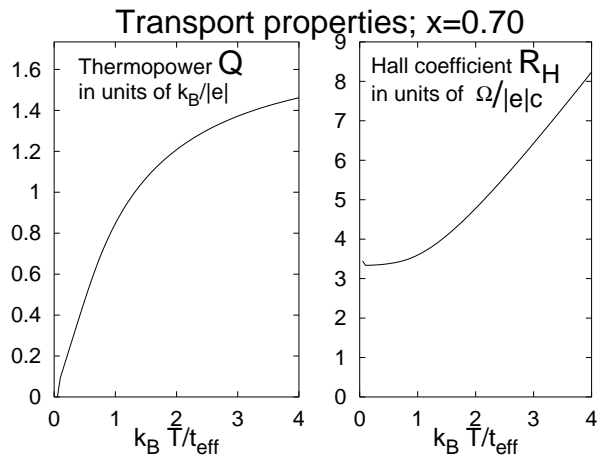}}
\vskip -2mm
\caption{Simple-minded transport theory for nondegenerate
spin-1/2 Fermi gas on the triangular lattice.
The thermopower is plotted in ``classical'' units of $k_B/|e|$,
while the units for the Hall coefficient contain the
three-dimensional volume per Co atom $\Omega$.
}
\label{QRHx70}
\end{figure}

b) The Hall coefficient for the triangular lattice band
structure has an unusual non-saturating increase with the
temperature for $T \gtrsim \epsilon_F$ as observed experimentally
in Ref.~\onlinecite{WangRH}.
The limiting high temperature behavior is
\begin{equation}
R_H = \frac{\Omega}{qc \,\rho(2-\rho)} \, \frac{k_B T}{t_{\rm eff}}~,
\label{RH:highT}
\end{equation}
where $\Omega$ is the three-dimensional volume per Co.
The full temperature dependence is shown in Fig.~\ref{QRHx70};
the high temperature trend sets in for $T \approx 1$ to $2\, 
t_{\rm eff}$.
Possibility of this unusual behavior was predicted in
Ref.~\onlinecite{KumarShastry} from the high temperature expansion
for the $tJ$ model (the doping dependence of the proportionality
coefficient is somewhat different here).
We remark that this unusual behavior is the consequence
of the triangular lattice band structure only, and its origin
can be traced to the presence of three-site hopping loops
as detailed in Appendix~\ref{app:transport}.
Correlation effects per se are needed only to reduce
$t_{\rm eff}$ below the experimental temperatures.

c) Pauli susceptibility per Co site for $T \gtrsim \epsilon_F$ becomes
\begin{equation}
\chi = \frac{\mu_B^2}{T} \rho (1-\rho/2) ~.
\end{equation}
Note that this has a Curie-like behavior, but is somewhat
smaller---by a factor of $1-\rho/2$---from the case of
completely free spins.

3) The kinetic energy renormalizations are strongest
(for fixed repulsion strength) near the commensurate $x=1/3,\, 2/3$,
and weakest near $x=1/2$.  This is because the system finds
it easiest to order, even if only locally, near the commensurate
filling, while away from commensuration much of the nearest neighbor
repulsion energy cannot be avoided in any case.

Charge frustration may also be relevant for the experimental
``charging'' curve of Ref.~\onlinecite{Chou}.
The observed plateaus at $x=1/3,\, 2/3$ remind one of the magnetization
plateaus in the frustrated triangular lattice Ising model
(related to the lattice gas with nearest neighbor repulsion as
mentioned in Sec.~\ref{sec:PsiJG}).  
Note that the bandwidth observed by heat capacity\cite{Chou,Ando} 
and by ARPES\cite{Has} is proportional to $\chi$ and has contributions 
from both $g_t t$ and $g_J J$ (see Eq.~\ref{selfcons_chi}).  
On the other hand, electromagnetic response couples only to $t$ 
so that the Drude weight observable from infrared reflectivity and
the superfluid density (observable via the London penetration depth 
in the case of superconductors) are directly proportional to $g_t$.  
These will provide a more sensitive test of
the predicted dip in $g_t$ near $x=1/3$ and $x=2/3$ 
as shown in Fig.~\ref{fig:gthop}.  For
example, it will be interesting to compare the Drude weight for 
$x = 1/3$ and $x=0.5$ samples.

4) The spin physics near $x=2/3$ is expected to be highly degenerate
and complicated, and will manifest itself below the energy scale
$t_{\rm eff}$.  In particular, the above transport pictures
will likely be modified below this scale.

5) Near $x=1/3$, whether the system prefers uniform or charge-ordered
state, the correlated liquid can have further RVB superconducting
instabilities.  The suppression of the charge mobility serves to
enhance and may even resurrect the superconductivity
under a small superconducting dome around $x=1/3$.
Experiments\cite{Schaak} observe the disappearance of the
superconductivity below $x=0.26$.  However, the strongest RVB
superconductivity is expected at much lower doping, and the search
should be pursued more vigorously towards $x=0$,
if that is chemically possible.

\acknowledgements
The authors are grateful to A. Vishwanath, C. Honerkamp, and T. Senthil
for many useful discussions, and to N. P. Ong for making experimental
results available before publication.
This work was supported by the National Science Foundation under
grants DMR--0201069 and DMR-0213282.
OIM also wants to thank his family for support during his stay in
Ukraine.

\appendix
\section{Details of Eq.~(\ref{DeltaBCS})}
\label{app:DeltaBCS}
Eq.~(\ref{DeltaBCS}) can be understood by examining the
self-consistency conditions Eq.~(\ref{selfcons_Delta}).
Specializing for the $d+id$ ansatz, we have
\begin{eqnarray*}
1 &=& \frac{3J_{\rm eff}}{8}
\frac{1}{N_{\rm latt}} \sum_{\bm k}
\frac{ f_{d+id}({\bm k}) }
     { \sqrt{\xi_{\bm k}^2 + \Delta_{\bm k}^2 } }~. \\
f_{d+id}({\bm k}) &\equiv&
2 \cos{\bm k}\cdot{\bm e}_1
\left(\cos{\bm k}\cdot{\bm e}_1 - \frac{1}{2}\cos{\bm k}\cdot{\bm e}_2
      - \frac{1}{2}\cos{\bm k}\cdot{\bm e}_3 \right) ~.
\end{eqnarray*}
Here $N_{\rm latt}$ is the number of lattice sites;
$\xi_{\bm k}=\epsilon_{\bm k}-\mu$ with
$\epsilon_{\bm k} = -2\chi ( \cos{\bm k}\cdot{\bm e}_1
+ \cos{\bm k}\cdot{\bm e}_2 + \cos{\bm k}\cdot{\bm e}_3 )$.

For weak superconductivity
$\Delta \ll J_{\rm eff} \lesssim t_{\rm eff}$,
following a BCS-like analysis, 
we obtain the following approximate formula
\begin{equation}
\Delta = A \; \chi \; 
\exp\left( -\frac{4}{3\nu_0(x) \overline{f_{d+id}}(x)}
            \frac{\chi}{J_{\rm eff}} \right) ~.
\label{apprTc}
\end{equation}
$A$ is an order one numerical constant,
$\nu_0(x)$ is the triangular lattice hopping density of states per site
(not including spin) at the Fermi energy corresponding to doping $x$,
and $\overline{f_{d+id}}$ is the $d+id$ wave factor averaged over the 
Fermi surface.  The scale $\chi$ in front of the exponential corresponds
to the energy cutoff being roughly the Fermi energy, since
the pairing is over the full Fermi volume.
For small $J_{\rm eff}$ we see from Eq.~(\ref{selfcons_chi}) that 
$\chi$ can be replaced by $t_{\rm eff}$ in (\ref{apprTc}), 
yielding Eq.~(\ref{DeltaBCS}).
Similar expression is obtained for the meanfield $T_c$.

\begin{figure}
\centerline{\includegraphics[width=\columnwidth]{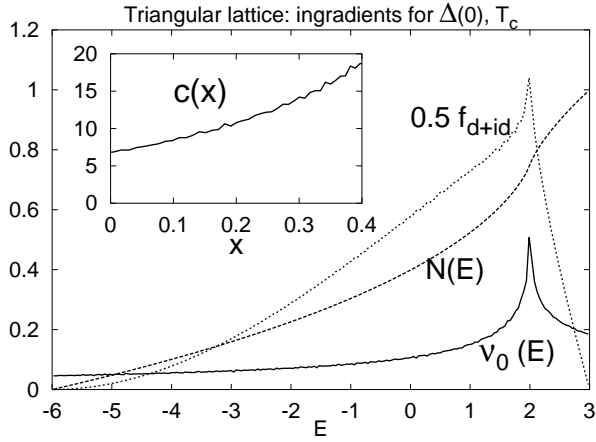}}
\vskip -2mm
\caption{Density of states per site $\nu_0(E)$ and the integrated
DOS $N(E)$ for the triangular lattice band structure 
(unit hopping amplitude, no spin).
We also show the data for the average $d+id$ wave factor
$\overline{f_{d+id}}$ at the corresponding energy cut;
we plot $0.5 \overline{f_{d+id}}$ to fit into the same vertical scale.
To obtain the values corresponding to the particular doping $x$,
we first locate $E$ such that $N(E) = (1-x)/2$.
In this manner, we obtain $c(x)= 4 /(3\nu_0 \overline{f_{d+id}})$
plotted for the relevant range $0<x<0.4$ in the inset.
}
\label{dosinfo}
\end{figure}

The necessary data is shown in Fig.~\ref{dosinfo}.
The coefficient $c(x)$ depends rather weakly on $x$,
and the main effect on the $\Delta$ and $T_c$ is
from the doping dependence of $t_{\rm eff}$.
The above approximate formula agrees fairly well with the
actual meanfield calculations performed in the main text.

\section{Transport for $T \lesssim \epsilon_F$}
\label{app:transport}
In this appendix, we summarize the simple Fermi liquid
transport theory that was used to obtain Fig.~\ref{QRHx70}
and Eqs.~\ref{Q:highT},\ref{RH:highT}.
The main formulae can be found in standard texts.~\cite{AM,Ziman}.

a) The thermopower is given by
\begin{equation}
Q = \frac{{\tt L}_{12}}{\sigma_{xx}} ~.
\end{equation}
The kinetic coefficients are given by the integrals
over the Brillouin zone
\begin{eqnarray*}
{\tt L}_{12} &=& q \tau
\int \frac{d^3 {\bm k}}{4 \pi^3}
\left(-\frac{\partial f}{\partial \epsilon} \right)
v_x({\bm k}) v_x({\bm k}) \frac{\epsilon({\bm k})-\mu}{T} ~, \\
\sigma_{xx} &=& q^2 \tau
\int \frac{d^3 {\bm k}}{4 \pi^3}
\left(-\frac{\partial f}{\partial \epsilon} \right)
v_x({\bm k}) v_x({\bm k}) ~.
\end{eqnarray*}
Here, $f(\epsilon)=1/(e^{\epsilon-\mu}+1)$ is the Fermi distribution.
We cite the more familiar three-dimensional expressions.
When applying to \nacoo, we specialize to the layered
triangular lattice by assuming no dispersion in the
$\hat{\bm z}$ direction.  The full temperature dependence for
$x=0.70$ is shown in Fig.~\ref{QRHx70},
and the limiting high-temperature behavior is given
in Eq.~(\ref{Q:highT}).

b) In weak magnetic fields $\omega_c \tau \ll 1$, the
Hall coefficient is given by
\begin{equation}
R_H = \frac{\sigma_H}{\sigma_{xx} \sigma_{yy}} ~.
\end{equation}
$\sigma_{xx}$, $\sigma_{yy}$ are the static zero field
conductivities given earlier, while
\begin{equation}
\sigma_H = \frac{q^3 \tau^2}{c}
\int \frac{d^3 {\bm k}}{4 \pi^3}
\left(-\frac{\partial f}{\partial \epsilon} \right)
v_x(\bm k) [M_{yy}^{-1} v_x({\bm k}) - M_{yx}^{-1} v_y({\bm k})] ~.
\label{sigmaH}
\end{equation}
In the last equation,
$M_{\alpha\beta}^{-1}({\bm k}) =
\frac{\partial^2 \epsilon}{\partial k_{\alpha} \partial k_{\beta}}$
is the inverse mass tensor.

The high temperature behavior for the layered triangular lattice
is given by Eq.~(\ref{RH:highT}).
The origin of this non-saturating increase with temperature lies in the
presence of triangular hopping loops.
Indeed, consider the above semi-classical expression for $\sigma_H$ at
high temperature, and translate it from the momentum space back to the
real space assuming a general hopping problem $t_{RR'}$ on a
Bravais lattice.  The result reads
\begin{equation}
\sigma_H = \frac{q^3 \tau^2}{c}
\left(-\frac{\partial f}{\partial \epsilon} \right)
\frac{2}{\Omega}
\sum_{{\bm R}_1, {\bm R}_2} t_{01} t_{12} t_{20} \, R_{1x} R_{2y}
({\bm R_1} \times {\bm R_2})_z ~.
\end{equation}
Here, $-\frac{\partial f}{\partial \epsilon} \approx \rho(2-\rho)/(4T)$,
and also enters $\sigma_{xx}, \sigma_{yy}$; $\Omega$ is the volume
of the unit cell.  The lattice hopping problem is input through the real
space sum over possible hops out of the origin:
${\bm R}_1 \equiv {\bm R}_{01}$, ${\bm R}_2 \equiv {\bm R}_{02}$.
For each triangle specified by an unordered triple of vertices
${\bm 0}, {\bm R}_1, {\bm R}_2$, the clockwise $0\to 1 \to 2 \to 0$
and anticlockwise $0 \to 2 \to 1 \to 0$ contributions add to
$({\bm R}_1 \times {\bm R}_2)_z^2$, i.e., a quantity of definite
sign.  The effect is of course strongest for the triangular
lattice.


\begin{thebibliography}{10}

\bibitem{Takada}
K. Takada\etal, Nature {\bf 422}, 53 (2003).

\bibitem{Foo}
M. L. Foo\etal, cond-mat/0304464;
B. Lorenz\etal, cond-mat/0304537; F. Rivadulla\etal, cond-mat/0304455.

\bibitem{Schaak}
R.~E.~Schaak, T.~Klimczuk, M.~L.~Foo, and R.~J.~Cava, cond-mat/0305450.

\bibitem{Chou}
F. C. Chou, J. H. Cho, P. A. Lee, E. T. Abel, K. Matan, and Y. S. Lee,
cond-mat/0306659


\bibitem{Terasaki}
I. Terasaki, Y. Sasago, and K. Uchinokura,
Phys. Rev. B {\bf 56}, R12685 (1997).

\bibitem{Ando}
Y. Ando, N. Miyamoto, K. Segawa, T. Kawata, and I. Terasaki,
Phys. Rev. B {\bf 60}, 10580 (1999).

\bibitem{Koshibae}
W. Koshibae, K. Tsutsui, and S. Maekawa,
Phys. Rev. B {\bf 62}, 6869 (2000).

\bibitem{Ray}
R. Ray, A. Ghoshray, K. Ghoshray, and S. Nakamura,
Phys. Rev. B {\bf 59}, 9454 (1999).

\bibitem{WangQ}
Y.~Wang, N.~S.~Rogado, R.~J.~Cava, and N.~P.~Ong,
Nature (2003).

\bibitem{WangRH}
Y.~Wang, N.~S.~Rogado, R.~J.~Cava, and N.~P.~Ong,
cond-mat/0305455.

\bibitem{Has}
M.~Z.~Hasan\etal, cond-mat/0308438.

\bibitem{Gav}
G.~L.~Gavilano\etal, cond-mat/0308383.

\bibitem{Lem}
O.~P.~Lemmens\etal, cond-mat/0309186.

\bibitem{Bru}
M.~Bruhwiler, B.~Batlogg, S.~M.~Kazakov and J.~Karpinski,
cond-mat/0309311.

\bibitem{Singh}
D. J. Singh, Phys. Rev. B {\bf 61}, 13397 (2000).

\bibitem{Baskaran2}
G.~Baskaran, cond-mat/0306569.

\bibitem{Baskaran1}
G.~Baskaran, cond-mat/0303649.

\bibitem{KumarShastry}
B.~Kumar and B.~S.~Shastry, cond-mat/0304210.

\bibitem{WangLeeLee}
Q.-H.~Wang, D.-H.~Lee, and P.~A.~Lee, cond-mat/0304377.

\bibitem{Vollhardt}
D. Vollhardt, Rev. Mod. Phys. {\bf 56}, 99 (1984).

\bibitem{FCZhang}
F. C. Zhang, C. Gros, T. M. Rice, and H. Shiba,
Supercond. Sci. Technol. {\bf 1}, 36 (1988).

\bibitem{Gros}
C. Gros, Annals Phys. (NY) {\bf 189}, 53 (1989);
C. Gros, Phys. Rev. B {\bf 38}, 931 (1988).

\bibitem{MEFisher}
M. E. Fisher, Rep. Prog. Phys. {\bf 30}, 615 (1967), and references
therein;
B. D. Metcalf, Phys. Lett. {\bf 45} A, 1 (1973);
M. Schick, J. S. Walker, and M. Wortis, Phys. Rev. B {\bf 16},
2205 (1977).

\bibitem{Baxter}
R. J. Baxter, J. Phys. A {\bf 13}, L61 (1981).

\bibitem{Ceperley}
D. M. Ceperley, G. V. Chester and M. H. Kalos,
Phys. Rev. B {\bf 16}, 3081 (1977).


\bibitem{Ogata}
M.~Ogata, cond-mat/0304405.

\bibitem{Honerkamp}
C.~Honerkamp, cond-mat/0304460.

\bibitem{AM}
N. Ashcroft and N. Mermin, Solid State Physics.

\bibitem{Ziman}
J. M. Ziman, Electrons and Phonons, Oxford, 1960.







\end{thebibliography}
\end{document}